\documentclass[a4paper,12pt]{article}
\usepackage{amssymb}
\usepackage{latexsym}
\usepackage[dvips]{graphicx}
\usepackage{cite}

\usepackage{ae} 
\usepackage[T1]{fontenc}
\usepackage[ansinew]{inputenc}
\usepackage{amsmath}
\usepackage{graphicx}
\usepackage{color}

\newcommand{\V}{\mathcal{V}}

\begin{document}
\begin{center}
{\LARGE{\bf Axially  Symmetric Cosmological Mesonic Stiff Fluid Models in Lyra's Geometry}}\\[1em]
\large{\bf{Ragab M. Gad}\footnote{Email Address: ragab2gad@hotmail.com}}\\
\normalsize {Mathematics Department, Faculty of Science,}\\
\normalsize  {Minia University, 61915 El-Minia,  EGYPT.}
\end{center}

\begin{abstract}
In this paper, we obtained a new class of axially symmetric cosmological mesonic stiff fluid models in the context of Lyra's geometry. Expressions for the energy, pressure and the massless scalar field are derived by considering the time dependent displacement field. We found that the mesonic scalar field depends on only $t$ coordinate. Some physical properties of the obtained models are discussed.
\end{abstract}

\setcounter{equation}{0}
\section{Introduction}
After Einstein proposed his theory of general relativity  he
succeeded in geometrizing the phenomena of gravitation by expressing
gravitational in terms of metric tensor $g_{ij}$. This idea of geometrizing
gravitation inspired many physicists to generalize the theory in
order to incorporate electrodynamics as a purely geometrical
construct. One of the first attempts in this direction was made in
1918 by Weyl \cite{W18} who suggested a theory based on a
generalization of Riemannian geometry, by formulating a new kind of
gauge theory involving metric tensor, to geometrize gravitation and
electromagnetism. This theory was criticized due to
non-integrability of length of vector under parallel displacement
\footnote{In the theory of general relativity if a vector undergoes
a parallel displacement, its direction may change, but not its
length. While in Weyl's geometry not only the direction but the
length may change and depends on the path between two points, which
show that the length is not integrable.}. Also as pointed out by
Einstein that this theory implies that frequency of spectral lines
emitted by atoms would not remain constant but would depend on their
past histories, which is in contradiction to observed uniformity of
their properties \cite{SD71}.
 Nevertheless, completely
apart from these criticisms, Weyl's geometry provides an interesting
example of non-Riemannian connections. Folland \cite{F70} gave a
global formulation of Weyl manifolds clarifying considerably many of
Weyl's basic ideas thereby.
\par
In 1951 Lyra \cite{L51} proposed a modification of Riemannian geometry
by introducing a gauge function into the structure less manifold
which is in close resemblance to Weyl's geometry \cite{S52}. This
modification was to overcome the problems appeared in Weyl's
geometry and is more in keeping with the spirit of Einstein's
principle of geometrization, since both the scalar and tensor fields
have more or less intrinsic geometrical significance. In this way
Riemannian geometry was given a new modification and the modified
geometry was named as Lyra's geometry. For physical motivation of
Lyra's geometry we refer to the literature \cite{S60}-\cite{H73}.
However, in contrast to Weyl's geometry, in Lyra's geometry the
connection is metric preserving and length transfers are integrable
as in Riemannian geometry. Subsequently, Sen \cite{Sen57} and Sen
and Dunn \cite{SD71} proposed a new scalar tensor theory of
gravitation. They constructed an analog of Einstein field equation
based in Lyra's geometry, see equation (\ref{FE}). Sen \cite{Sen57}
found that static model with finite density in Lyra's geometry is
similar to the static Einstein model, but a signification
differences was that the model exhibited red shift. Halford
\cite{H70} pointed out that the constant displacement vector field
$\phi_{i}$ in Lyra's geometry plays the role of a cosmological
constant in the normal general relativistic treatment. Halford
\cite{H72} showed that the scalar tensor treatment based in Lyra's
geometry predicts the same effects, within observational limits, as
in Einstein theory.  Several attempts have been made to cast the
scalar tensor theory of gravitation in wider geometrical context
\cite{S1}.
\par
Many Authors \cite{7} have studied cosmological models based on
Lyra's geometry with a constant displacement field vector in the time-direction.
 Singh and his collaborators \cite{8}
have studied Bianchi type I, III, Kantowski-Sachs and new class of
models with a time dependent displacement field. They have made a
comparative study of Robertson-Walker models with a constant
deceleration parameter in Einstein's theory with a cosmological
terms and in the cosmological theory based on Lyra's geometry.
\par
Recently, several authors \cite{P09}- \cite{S08} studied cosmological models based on Lyra's geometry in various contexts. With these motivations, in this paper, we obtained exact solutions of the field equations for mesonic stiff fluid models in axially symmetric space-times within the frame work of Lyra's geometry for  time varying displacement  field vector.
\par
Axially symmetric cosmological models have been studied in both Riemannian and Lyra geometries. In context of general relativity theory, by adopting the comoving coordinate system, these models with string dust cloud source are studied by Bhattacharaya and Karade \cite{BK93}. They shown that some of these models are singular free even at an initial epoch. In the context of Lyra's geometry these models are studied in the presence of cosmic  string source and thick domain walls \cite{RR06} and in the presence of perfect fluid distribution \cite{RV09}.
\par
This paper is organized as follows: The metric and field equations are presented in section 2. Section 3  deals with   solving the field equations. Finally, in section 4, concluding remarks are given.

\setcounter{equation}{0}
\section{Fundamental Concepts and Field Equations}
Consider the axially symmetric metric \cite{BK93} in the form
\begin{equation}\label{met}
ds^2 = dt^2 -A^2(t)(d\chi^2 +f^2(\chi) d\phi^2) -B^2(t)dz^2,
\end{equation}
with the convention $x^0=t$, $x^1=\chi$, $x^2=\phi$, $x^3=z$ and $A$ and $B$ are functions of $t$ only while $f$ is a function of the coordinate $\chi$ only.\\

The volume element of the model (\ref{met}) is given by
\begin{equation}\label{vol}
\V = \sqrt{-g}= A^2fB
\end{equation}
The four-acceleration vector, the rotation, the expansion scalar and the shear
scalar characterizing the four velocity vector field, $u^a$,
respectively, have the usual definitions as given by Raychaudhuri
\cite{R79}
\begin{equation}\label{Kin}
\begin{array}{ccc}
\dot{u}_i & = & u_{i;j}u^j ,\\
\omega_{ij} & = & u_{[i;j]}+\dot{u}_{[i}u_{j]},\\
\Theta & =  & u^i_{;i},\\
\sigma^2 & =& \frac{1}{2}\sigma_{ij}\sigma^{ij},
\end{array}
\end{equation}
where
$$
\sigma_{ij} =u_{(i;j)}+\dot{u}_{(i}u_{j)} -\frac{1}{3}\Theta
(g_{ij}+u_iu_j).
$$
In view of the metric (\ref{met}), the four-acceleration vector, the rotation, the
expansion scalar and the shear scalar given by (\ref{Kin})can be
written in a comoving coordinates system as

\begin{equation}\label{Kin-v}
\begin{array}{ccc}
\dot{u}_i & =& 0,\\
\omega_{ij} & = & 0,\\
\Theta & = & \frac{2\dot{A}}{A}+\frac{\dot{B}}{B},\\
\sigma^2 & = & \frac{1}{9}\big(11\big(\frac{\dot{A}}{A}\big)^2+5\big(\frac{\dot{B}}{B}\big)^2 + \frac{2\dot{A}\dot{B}}{AB}\big).
\end{array}
\end{equation}
The non vanishing components of the shear tensor$\sigma_{ij}$ are
\begin{equation}\label{comp}
\begin{array}{ccc}
\sigma_{11} & =  & A(\frac{1}{3}\Theta A-\dot{A}) ,\\
\sigma_{22} & = & Af^2(\frac{1}{3}\Theta A-\dot{A}) ,\\
 \sigma_{33}& = & B(\frac{1}{3}\Theta B-\dot{B}),\\
 \sigma_{44}& =  & -\frac{2}{3}\Theta.

\end{array}
\end{equation}

The field equations in normal gauge for Lyra's geometry as obtained
by Sen \cite{Sen57} (in gravitational units $c=8\pi G=1)$ read as
\begin{equation}\label{FE}
R_{ij}-\frac{1}{2}Rg_{ij}=-T_{ij}-\frac{3}{2}\phi_{i}\phi_{j}+
\frac{3}{4}g_{ij}\phi_{\alpha}\phi^{\alpha},
\end{equation}
the left hand side is the usual Einstein tensor, whereas  $\phi_{i}$ is a time-like displacement field vector defined by
$$
\phi_{i}=(0,0,0,\lambda(t)),
$$
and $T_{ij}$ is the energy momentum tensor corresponding to perfect
fluid and massless mesonic scalar field and is given by

\begin{equation}\label{EM}
T_{ij}=(\rho +p)u_{i}u_{j}-pg_{ij} + V_{,i}V_{,j}
-\frac{1}{2}g_{ij}V_{,k}V^{,k}.
\end{equation}
Here $p$ is the pressure, $\rho$ the energy density and $u_{i}$ the
four velocity vector satisfying the relation in co-moving coordinate
system
$$
g_{ij}u^iu^j=1, \qquad u^i=u_{i}=(1,0,0,0).
$$
However, $V$ is the massless scalar field and we assume it to be a
function of $t$ and $\chi$ coordinates. The Scalar field $V$ is
governed by the Klein-Gordan wave equation
$$
g^{ij}V_{;ij}=0.
$$

For the line element (\ref{met}), the field equations (\ref{FE})
with equation (\ref{EM})  lead to the following system of equations
\begin{equation}\label{1}
\frac{\ddot{A}}{A} + \frac{\ddot{B}}{B} + \frac{\dot{A}\dot{B}}{AB}
+ \frac{3}{4}\lambda^2 = -(p+\frac{1}{2}\dot{V}^2)-\frac{V^{\prime
2}}{2A^2},
\end{equation}
\begin{equation}\label{2}
\frac{\ddot{A}}{A} + \frac{\ddot{B}}{B} + \frac{\dot{A}\dot{B}}{AB}
+ \frac{3}{4}\lambda^2 = -(p+\frac{1}{2}\dot{V}^2)+\frac{V^{\prime
2}}{2A^2},
\end{equation}
\begin{equation}\label{3}
\frac{2\ddot{A}}{A} + (\frac{\dot{A}}{A})^2 -
\frac{f^{\prime\prime}}{fA^2} +\frac{3}{4}\lambda^2 =
-(p+\frac{1}{2}\dot{V}^2)+\frac{V^{\prime 2}}{2A^2},
\end{equation}
\begin{equation}\label{4}
 (\frac{\dot{A}}{A})^2 +\frac{2\dot{A}\dot{B}}{AB}
 - \frac{f^{\prime\prime}}{fA^2}-\frac{3}{4}\lambda^2 =
 (\rho +\frac{1}{2}\dot{V}^2)+\frac{V^{\prime 2}}{2A^2},
\end{equation}
\begin{equation}\label{5}
 \dot{\rho}+(\rho + p)(\frac{2\dot{A}}{A}+\frac{\dot{B}}{B})=0,
\end{equation}
\begin{equation}\label{6}
\ddot{V}-\frac{1}{A^2}V^{\prime\prime}+(\frac{2\dot{A}}{A}
+\frac{\dot{B}}{B})\dot{V} -\frac{ff^{\prime}}{f^2A^2}V^{\prime}=0.
\end{equation}
Here the over heat dot denotes differentiation with respect to $t$ and
over head prime denotes differentiation with respect to $\chi$.\\
From equations (\ref{1}) and (\ref{2}), we get
\begin{equation}\label{7}
V^{\prime}= 0.
\end{equation}
Consequently, the mesonic scalar field does not exit in the direction of $\chi$.\\

The functional dependence of the metric together with equations (\ref{2})
and (\ref{3}), using equation (\ref{7}), imply that
\begin{equation}\label{8}
\frac{f^{\prime\prime}}{f}=k^2, \qquad k^2= \text{constant}.
\end{equation}
If $k=0$, then the solution of this differential equation is
$f(\chi) =k_{1}\chi + k_{2}$, $k_{1}$ and $k_{2}$ are constants of
integration. Without loss of generality, we choose $k_{1} =1$ and
$k_{2}=0$. Thus we shall have
\begin{equation} \label{9}
f(\chi)= \chi.
\end{equation}

\par
 In the case $f(\chi)= \chi$ the field equations (\ref{1})-(\ref{6}), using
equation (\ref{7}), reduce to
\begin{equation}\label{11}
\frac{\ddot{A}}{A} + \frac{\ddot{B}}{B} + \frac{\dot{A}\dot{B}}{AB}
+ \frac{3}{4}\lambda^2 = -(p+\frac{1}{2}\dot{V}^2),
\end{equation}
\begin{equation}\label{12}
\frac{2\ddot{A}}{A} + (\frac{\dot{A}}{A})^2  +\frac{3}{4}\lambda^2 =
-(p+\frac{1}{2}\dot{V}^2),
\end{equation}
\begin{equation}\label{13}
 (\frac{\dot{A}}{A})^2 +\frac{2\dot{A}\dot{B}}{AB}
 -\frac{3}{4}\lambda^2 =
 (\rho +\frac{1}{2}\dot{V}^2),
\end{equation}
\begin{equation}\label{14}
 \dot{\rho}+(\rho + p)(\frac{2\dot{A}}{A}+\frac{\dot{B}}{B})=0,
\end{equation}
\begin{equation}\label{15}
\ddot{V}+(\frac{2\dot{A}}{A} +\frac{\dot{B}}{B})\dot{V}=0.
\end{equation}

We can easily find from (\ref{15}) that
\begin{equation}\label{V}
\dot{V} =\frac{n}{A^2B},
\end{equation}
where $n (\neq 0)$ is a constant of integration.

From equations (\ref{11}) and (\ref{12}), we get
\begin{equation}\label{16}
 \frac{\ddot{B}}{B} + \frac{\dot{A}\dot{B}}{AB}=\frac{\ddot{A}}{A} +
 (\frac{\dot{A}}{A})^2.
\end{equation}
We assume $A$ to be some arbitrary  function of $B$, say
\begin{equation}\label{16-}
A=\psi(B).
\end{equation}
So equation (\ref{16}) becomes
\begin{equation}\label{17}
\big(\frac{\psi_{B}}{\psi}-\frac{1}{B}\big)\ddot{B} +
\big[\frac{\psi_{BB}}{\psi}+
(\frac{\psi_{B}}{\psi})^2-\frac{\psi_{B}}{B\psi}\big]\dot{B}^2=0,
\end{equation}
where $\psi_{A}=\frac{d\psi}{dA}$. Equation (\ref{17}) results in
the following possibilities.\\
{\bf{(i-1)}}\\
\begin{equation}\label{18}
\frac{\psi_{B}}{\psi}-\frac{1}{B}=0, \quad \text{and} \qquad
 \frac{\psi_{BB}}{\psi}+
(\frac{\psi_{B}}{\psi})^2-\frac{\psi_{B}}{B\psi}=0,
\end{equation}
{\bf{(i-2)}}\\
\begin{equation}\label{19}
\ddot{B}=0, \quad \text{and} \qquad
 \frac{\psi_{BB}}{\psi}+
(\frac{\psi_{B}}{\psi})^2-\frac{\psi_{B}}{B\psi}=0,
\end{equation}
{\bf{(i-3)}}\\
\begin{equation}\label{20}
\dot{B}=0.
\end{equation}

\setcounter{equation}{0}
\section{Solutions of Field Equations}
 We have only five highly non-linear field equations
(\ref{11})-(\ref{15}) in sex unknowns , $A, B, p,
\rho, V$ and $\lambda$. In order to obtain its exact solution, we
assume one more physically reasonable condition amongst these
variables. We consider here the effective "stiff fluid"
distribution, that is, a perfect fluid with the equation of state:
\begin{equation}\label{21}
p=\rho.
\end{equation}
The equation of state (\ref{21}) was apparently first proposed by
Zeldovich \cite{Z62}. It should have applied in the early Universe,
the justification being the observation that with (\ref{21}) the
velocity of sound equals the velocity of light, so no material in
this Universe could be more stiff.\\
Using the condition (\ref{21}) in equation (\ref{14}) and by
integrating,  we get
\begin{equation}\label{PV}
p=\rho = \frac{m}{A^4B^2},
\end{equation}
where $m (\neq 0)$ is a constant of integration.

\underline{\bf{Case (i-1):}}\\
From equation (\ref{18}), using the first equation in the second
equation, then  by integrating, we get
\begin{equation}\label{22}
\psi = c_{1}B+c_{2},
\end{equation}
where $c_{1} (\neq 0)$ and $c_{2}$ are  integration constants.
Using this result in the first equation in (\ref{18}), we get
$c_2=0$. So that equation (\ref{22}) becomes
\begin{equation}\label{23}
\psi = c_{1}B.
\end{equation}
Using this result in equation (\ref{17}), we have
\begin{equation}\label{24}
A = c_{1}B.
\end{equation}
Now using this equation and the condition  (\ref{21}), in equations
(\ref{11})-(\ref{13}), we get

\begin{equation}\label{25}
\frac{2\ddot{B}}{B} + (\frac{\dot{B}}{B})^2  +\frac{3}{4}\lambda^2 =
-\frac{3}{2}\rho,
\end{equation}
\begin{equation}\label{26}
 3(\frac{\dot{B}}{B})^2  -\frac{3}{4}\lambda^2 =
 \frac{3}{2}\rho.
\end{equation}
These equations yield
\begin{equation}\label{27}
B=(at+b)^{\frac{1}{3}},
\end{equation}
where $a(\neq 0)$ and $b$ are constants of integration.\\
According to equations (\ref{24}) and (\ref{27}) the line element (\ref{met}) can be written in the following form
\begin{equation}\label{28}
ds^2 = dt^2 -(at+b)^{\frac{2}{3}}(c_1^2 d\chi^2 + c_1^2 \chi^2
d\phi^2+  dz^2),
\end{equation}

{\bf{Physical properties of the model}}\\

 Using equations (\ref{24}) and (\ref{27}) in equations
 (\ref{13})-(\ref{15}), take into account (\ref{21}), the expressions for density $\rho$, pressure $p$,  massless scalar field $V$ and displacement field $\lambda$ are given by
 $$
 \rho = p= \frac{c}{(at+b)^2},  \qquad c=\frac{m}{c_{1}^4},
 $$
 which shows that $\rho$ and $p$ are not singular,
 $$
 V=\frac{n}{ac_1^2}\log (at+b) +c_{3},
 $$
 $$
 \lambda^2 = \frac{c_{4}}{(at+b)^2}, \qquad c_{4}=\frac{4a^2}{9}-\frac{2(2m+n^2)}{3c_1^4}.
 $$
It is observed, from equations (\ref{24}) and (\ref{27}) that $A(t)$ and $B(t)$ can be singular only for $t \rightarrow \infty$. Thus the line element (\ref{28}) is singular free even at $t=0$.\\

For the line element (\ref{28}), using equations (\ref{vol}), (\ref{Kin-v}) and (\ref{comp}), we have the following physical properties:\\
The volume element is
$$
\V =c_1^2 \chi (at+b).
$$
This equation shows that the volume increases as the time increases, that is, the model (\ref{28}) is expanding with time.\\
The expansion scalar, which determines the volume behavior of the fluid, is given by
$$
\Theta = \frac{a}{at+b}.
$$
The only non-vanishing component of the shear tensor, $\sigma_{ij}$, is
$$
\sigma_{44}=- \frac{2a}{3(at+b)}.
$$
Hence, the shear scalar $\sigma$ is given by
$$
\sigma^2 =2\big(\frac{a}{3(at+b)}\big)^2.
$$

Since $\lim_{t\rightarrow\infty} (\frac{\sigma}{\Theta})\neq 0$, then the model (\ref{28}) does not approach isotropy for large value of $t$. Also the model does not admit  acceleration and rotation, since $\dot{u}_i =0$ and $\omega_{ij}=0$ .

\underline{\bf{Case (i-2):}}\\

The first equation in (\ref{19}) gives
\begin{equation}\label{i-2-1}
B=b_{1}t+b_{2},
\end{equation}
where $b_1 (\neq 0)$ and $b_2$ are constants of integration. Using
this result in the second equation in (\ref{19}) and take into
account (\ref{16-}), we get
$$
A^2= b_{3}(b_{1}t^2+2b_2t)+b_4,
$$
where $b_3 (\neq 0)$ and $b_4$ are constants of integration. This
equation can be written in the form
\begin{equation}\label{i-2-2}
A=( b_{3}B^2+b_5)^{\frac{1}{2}},
\end{equation}
where the constant $b_5$ depends on the constants $b_1$ and $b_2$.\\
Using equations (\ref{i-2-1}) and (\ref{i-2-2}) in equations
(\ref{11})-(\ref{15}), take into account the condition $\rho = p$,
we get
$$
 \rho = p= \frac{m}{(b_{1}t+b_{2})^2}[b_3(b_{1}t+b_{2})^2 + b_5]^2,
 $$
 $$
 V=\frac{n}{b_3b_5}\log \frac{b_{1}t+b_{2}}{\sqrt{b_3(b_{1}t+b_{2})^2 + b_5}}+b_6,
 $$
 where $b_6$ is the constant of integration,
 $$
 \lambda^2 = \frac{2(2b_1^2b_3^2(b_{1}t+b_{2})^3-2m-n^2)}
 {3(b_3(b_{1}t+b_{2})^2+b_5)^2(b_{1}t+b_{2})^2} -\frac{8b_1^2b_3}{3(b_3(b_{1}t+b_{2})^2+b_5)}.
 $$
In this case the line element (\ref{met}) takes the following form
\begin{equation}\label{m-i-2}
ds^2 = dt^2 -(b_3(b_1t+b_2)^2+b_5)(d\chi^2 +\chi^2 d\phi^2)-
(b_1t+b_2)^2 dz^2.
\end{equation}

We shall now give the expression for kinematic quantities. A straightforward calculation leads to the expressions for element volume $\V$, expansion scalar $\Theta$ and shear tensor $\sigma_{ij}$ of  model (\ref{m-i-2}) are given, respectively, by

$$
\V = \chi(b_1t+b_2)(b_3(b_1t+b_2)^2+b_5),
$$
which shows that the model is expanding with time,
$$
\Theta = \frac{2b_1b_3(b_1t+b_2)}{(b_3(b_1t+b_2)^2+b_5)} + \frac{b_1}{(b_1t+b_2)},
$$
$$
\sigma_{11} = \frac{b_1(b_3(b_1t+b_2)^2+b_5)}{3(b_1t+b_2)}-\frac{b_1b_3(b_1t+b_2)}{3},
$$
$$
\sigma_{22} = \chi^2\sigma_{11},
$$
$$
\sigma_{33} = \frac{2b_1b_3(b_1t+b_2)^3}{b_3(b_1t+b_2)^2+b_5}-\frac{2b_1(b_1t+b_2)}{3},
$$
$$
\sigma_{44} = -\frac{4b_1b_3(b_1t+b_2)}{3(b_3(b_1t+b_2)^2+b_5)} - \frac{2b_1}{3(b_1t+b_2)},
$$
and other components of the shear tensor, $\sigma_{ij}$, being zero. Hence
$$
\sigma^2 = \frac{1}{9}\Big( 11\frac{b_1^2b_3^2(b_1t+b_2)^2}{(b_1t+b_2)^2+b_5)^2} +5 \big(\frac{b_1}{b_1t+b_2}\big)^2 + \frac{2b^2_1b_3}{b_3(b_1t+b_2)^2+b_5}\Big).
$$
Moreover, this model represents non-rotating and has vanishing acceleration.
Since $\lim_{t\rightarrow \infty}(\frac{\sigma}{\Theta}) \neq 0$, then the model (\ref{m-i-2}) do not approach isotropy for large value of $t$.

\underline{\bf{Case (i-3):}}\\

Equation (\ref{20}) can be easily integrated to give
\begin{equation}\label{i-3-1}
B= \ell,
\end{equation}
where $\ell (\neq 0)$ is the constant of integration. Using this
equation in equations (\ref{11})-(\ref{13}) and take into account
the condition (\ref{PV}),we have
\begin{equation}\label{i-3-2}
A=(\ell_{1}t+ \ell_2)^{\frac{1}{2}},
\end{equation}
where $\ell_1 (\neq 0)$ and $\ell_2$ are constants of integration.
Using equations (\ref{i-3-1}) and (\ref{i-3-2}) in equations
(\ref{11})-(\ref{15}), take into account the condition $\rho = p$,
we get
$$
 \rho = p= \frac{m}{\ell^2(\ell_{1}t+ \ell_2)^2},
 $$
 $$
 V=\frac{n}{\ell \ell_1}\log (\ell_{1}t+ \ell_2) +  \ell_3,
 $$
 where $\ell_3$ is the constant of integration,
 $$
 \lambda^2 = \frac{\ell_4}{(\ell_1 + \ell_2)^2}, \qquad \ell_4 = \big(3\ell_1^2 - \frac{2(2m+n^2)}{3\ell^2}.
 $$
In this case the line element (\ref{met}) can be written in the form
\begin{equation}\label{m-i-3}
ds^2 = dt^2 -(\ell_{1}t+\ell_2)(d\chi^2 +\chi^2 d\phi^2)- \ell^2
dz^2.
\end{equation}
From equation (\ref{i-3-2}), one can observed that $A(t)$ is singular only when $t \rightarrow \infty$. Consequently, the line element (\ref{m-i-3}) is singular free even $t=0$.\\
For the line element (\ref{m-i-3}), the volume element $\V$ and the kinematics properties (acceleration $\dot{u}_i$, rotation $\omega_{ij}$, expansion scalar $\Theta$, shear tensor $\sigma_{ij}$ and  shear scalar $\sigma$) respectively found to have the following expressions:
$$
\V = \ell \chi (\ell_1t+\ell_2),
$$
which shows that the model is expanding with time,
$$
\dot{u} =0,
$$
$$
\omega_{ij} =0,
$$
$$
\Theta = \frac{\ell_1}{\ell_1t+\ell_2},
$$

$$
\sigma_{11} = -\frac{\ell_1}{6},
$$
$$
\sigma_{22} = \chi^2 \sigma_{11},
$$
$$
\sigma_{33} = \frac{\ell\ell_{1}}{3(\ell_1t+\ell_2)},
$$
$$
\sigma_{44} = -\frac{2\ell_1}{3(\ell_1t+\ell_2)}.
$$
$$
\sigma^2 = \frac{11\ell_1^2}{36(\ell_1t+\ell_2)}.
$$
As in the above two cases, $\lim_{t\rightarrow \infty}(\frac{\sigma}{\Theta}) \neq 0$,  the model (\ref{m-i-3}) do not approach isotropy for large value of $t$.

\setcounter{equation}{0}
\section{Conclusions}
This paper deals with axially symmetric space-time in the presence of mesonic stiff fluid distribution within the framework of Lyra's geometry for time dependent displacement field. We have presented a new class of exact solutions of Einstein's field equations for this space-time. The obtained models represent shearing, non-rotating and expanding with time $t$. Moreover, these models are singular free even at the initial epoch  $t=0$ and have vanishing accelerations. For all models, we found also that $\lim_{t\rightarrow \infty}(\frac{\sigma}{\Theta}) \neq 0$, this means that they are not approach isotropy for large time $t$. We found also that the mesonic scalar field in axially symmetric space-time exists only in the t-direction.


\end{document}